\documentclass{article}
\usepackage{graphicx}
\usepackage{comment}
\usepackage{amssymb} 
\usepackage{amsmath}
\newtheorem{theorem}{Theorem}
\newtheorem{lemma}{Lemma}
\newtheorem{proposition}{Proposition}
\newtheorem{corollary}{Corollary}
\newtheorem{example}{Example}
\newtheorem{definition}{Definition}
\newtheorem{observation}{Observation}
\newtheorem{claim}{Claim}
\date{}

\title{On oblivious branching programs with bounded repetition that cannot efficiently compute CNFs of bounded treewidth}
\author{Igor Razgon\\ Department of Computer Science and Information Systems,\\ Birkbeck, University of London\\
        igor@dcs.bbk.ac.uk}
\begin{document}
\maketitle
\begin{abstract}
In this paper we study complexity of an extension of ordered binary decision diagrams ({\sc obdd}s)
called $c$-{\sc obdd}s on {\sc cnf}s of bounded (primal graph) treewidth.
In particular, we show that for each $k$ there is a class of {\sc cnf}s of treewidth
$k \geq 3$ for which the equivalent $c$-{\sc obdd}s are of size $\Omega(n^{k/(8c-4)})$.
Moreover, this lower bound holds if $c$-{\sc obdd} is non-deterministic and semantic. 
Our second result uses the above lower bound to separate the above model from sentential 
decision diagrams ({\sc sdd}s).
In order to obtain the lower bound, we use a structural graph parameter called matching width.
Our third result shows that matching width and pathwidth are linearly related. 
\end{abstract}

\section{Introduction}
Ordered Binary Decision Diagrams {\sc obdd}s is a famous representation of Boolean
functions being actively investigated from both applied and theoretical perspective.
The theoretical research, among other things, has resulted in many upper and lower
bounds of {\sc obdd} size realizing various classes of functions \cite{WegBook}.

One such an upper bound, established in \cite{VardiTWD} states that a {\sc cnf} of
treewidth $k$ of its primal graph can be represented by an {\sc obdd} of size $O(n^k)$. 
In terms of parameterized complexity, this is an {\sc xp} upper bound, that is the degree
of the polynomial depends on $k$. A natural open question is whether this upper bound
can be improved to an {\sc fpt} upper bound, i.e. one of the form $f(k)*n^c$, where $c$
is a universal constant. 

This question is of a particular interest in the area of knowledge compilation because
of the recent introduction of Sentential Decision Diagrams ({\sc sdd}s) \cite{SDD} 
for which an {\sc fpt} upper bound \emph{does} hold. {\sc sdd}s share
with {\sc obdd}s a number of nice properties and have a good potential to replace {\sc obdd}s
in applications. Yet {\sc obdd}-related machinery is much more developed (one reason for that is
that {\sc obdd}s
have been investigated for a much longer time) and hence it is interesting to say if this
gap between upper bounds can be significantly tightened by finding a better upper bound for 
{\sc obdd}s.

In \cite{RazgonKR14}, we answered this question negatively by demonstrating that for each
$k \geq 3$ there is a class of {\sc cnf}s of primal graph treewidth at most $k$ for which the size
of equivalent {\sc obdd}s is $\Omega(n^{k/4})$. In this paper, which can be considered
as a follow-up version of \cite{RazgonKR14}, our motivation is to see how far the {\sc obdd}
can be extended so that the above lower bound would hold for that extended model in a way
that the lower bound in \cite{RazgonKR14} would follow as a special case.
As a result, we extend {\sc obdd}s as follows. First, for an arbitrary (but fixed) 
constant $c$ we use $c$-{\sc obdd}s instead {\sc obdd}. That is, we allow each variable to occur at
most $c$ times along each computational path, however the occurrences are ordered as $c$
concatenated copies of the same fixed permutation (in this setting the {\sc obdd} is simply
$1$-{\sc obdd}). Second, we allow the model to be non-deterministic. Roughly speaking,
this means that instead of applying this restriction on a branching program, the restriction
is applied on a switching and rectifier network. Third, we allow this restriction to be \emph{semantic},
i.e to hold only for \emph{consistent} paths that do not contain opposite occurrences of the same
variables. The in-consistent paths are not constrained at all. We call the resulting model 
Nondeterministic Semantic $c$-{\sc obdd} and abbreviate it $c$-{\sc nsobdd}.
In particular, we show that for each fixed $k \geq 3$ there is a class of {\sc cnf}s 
(in fact, the same class as we used in \cite{RazgonKR14}) for which the smallest
$c$-{\sc nsobdd} is of size $\Omega(n^{k/(8c-4)})$. Clearly, the lower bound for {\sc obdd}s follows
if we substitute $c=1$. 

The above lower bound shows that $c$-{\sc nsobdd}s are inherently different from {\sc sdd}
with respect to representation of {\sc cnf} of bounded treewidth. Our second result shows that
this difference can, in fact, be turned into a (non-parameterized) separation.
In particular by, essentially, setting $k$ to $\log n$, we obtain a class of
{\sc cnf} that can be represented by polynomial size {\sc sdd}s but require
$c$-{\sc nsobdd} of quasipolynomial size.

Our third result is related to the way the main lower bound is obtained.
In particular, the {\sc cnf}s we consider for the sake of obtaining lower bounds,
correspond to undirected graphs. We introduce a graph parameter called \emph{matching width}
and show that the size of $c$-{\sc nsobdd} equivalent to the considered {\sc cnf}
is exponential in the matching width of the corresponding graph. 
Then we show that there are graphs for which the matching width is $\Omega(\log n)$
times larger than their treewidth. The lower bound readily follows from the combination 
of these results. The relationship between matching width and treewidth suggests
that the former is \emph{similar} to pathwdith. Our third result shows that this is
indeed true, that is pathwidth and matching width are linearly related. 

The last result might seem a little bit out of scope. The reason why we provide it in this
paper is that matching width has already been used several time to obtain lower bounds \cite{RazgonKR14,RazgonIPEC14,Bova14}.
So, it is interesting to see how it is connected to well known graph parameters.
To the best of our knowledge \cite{RazgonKR14} is the first paper where matching width for used
for lower bounds, so a follow-up version of \cite{RazgonKR14}, seems the natural place
for showing how matching width is connected to pathwidth. 

Let us overview the related work.
The $c$-{\sc obdd} have been considered in \cite{Krausekobdd} with exponential lower bound provided for several
functions. The $c$-{\sc obdd} model is known to be more powerful than the ordinary {\sc obdd}. In particular,
Theorem 7.2.2. of \cite{WegBook} provides a class of functions polynomial for $2$-{\sc obdd} and exponential
even for Free Binary Decision Diagrams ({\sc fbdd}) (that is, read-once branching programs).
Moreover, it is known that increse of $c$ adds computational power. In particular, it has been demonstrated
in \cite{BolSau} that for each $c \geq 2$ there is a class of functions computable by poly-size
$c$-{\sc obdd}s and requiring exponential size $c-1$-obdds. Interesting refinements of this hierarchy involving
width of branching programs have been proposed in \cite{Ablay,Khadiev15}. 

It is also known that non-determinism adds power to {\sc obdd}. In particular, Theorem 10.2.3. of \cite{WegBook}
demonstrates a class of functions that can be computed by poly-size non-deterministic {\sc obdd}s, yet require
exponential size {\sc fbdd}s. We are not aware of the any existing research \emph{specifically} on non-deterministic $c$-{\sc obdd}s.
They are obviously a special case of non-deterministic read $k$-times branching programs and hence exponential
lower bounds (e.g. \cite{readktimes}) apply to them. It is well known that semantic rather than syntactic restriction
adds a lot of power if the obliviousness requirement is dropped. In particular, \cite{Yukna} demonstrates a class 
of functions that can be computed by poly-size semantic non-deterministic read-once branching programs but require
exponential size if `semantic' is replaced by `syntactic'. In fact, no super-polynomial lower bound is known for the former.
We are not aware, however, if the semantic restriction adds any power to non-deterministic {\sc obdd}s.
The lower bound of \cite{RazgonKR14} has been generalized in \cite{RazgonIPEC14} to a different direction than the
one considered in this paper: namely the obliviousness was dropped. In particular, it has been shown that
the non FPT lower bound holds for non-deterministic read-once branching programs.

Matching width can be seen as a special case of maximum matching width introduced in \cite{VaThesis} when the underlying
tree is a caterpillar. It has been shown in \cite{VaThesis} that maximum matching width is linearly related to the treewidth.
The linear relationship between matching width and pathwdith, established in this paper, looks natural in this context.

The rest of the paper is structured as follows.
Section 2 introduces the necessary background. 
Section 3 states the lower bound on $c$-{\sc nsobdd}s along with the separation
from {\sc sdd}. The lower is proved in Section 4.
The proof of linear relationship between matching width and pathwidth is provided in Section 5.

\section{Preliminaries} \label{prelim}

In this paper by a \emph{set of literals} we mean one that does not
contain an occurrence of a variable and its negation.
For a set $S$ of literals we denote by $Var(S)$ the set of variables
whose literals occur in $S$. If $F$ is a Boolean function
or its representation by a {\sc cnf} or {\sc obdd}, we denote by $Var(F)$
the set of variables of $F$. A truth assignment to $Var(F)$ on which $F$
is true is called a \emph{satisfying assignment} of $F$. A set $S$ of literals
represents the truth assignment to $Var(S)$ where variables occurring
positively in $S$ (i.e. whose literals in $S$ are positive) are assigned with $true$
and the variables occurring negatively are assigned with $false$.

A \emph{non-deterministic branching program} $Z$
is a directed acyclic graph {\sc dag} with one root $rt$ and one leaf $lf$.
Some of the edges of $Z$ are labelled with literals of variables.
A path $P$ of $Z$ is \emph{consistent} 
if it does not have two edges labelled with opposite occurrences of the same
variable. This gives us possibility to define $A(P)$, the set of literals
labelling the edges of a consistent path $P$. A consistent root-leaf path of $Z$ is also called
a \emph{computational} path. The function $F$ computed by $Z$ is defined as follows.
Let $S$ be an assignment to the variables of $Z$. Then $S$ is a satisfying assignment
of $F$ if and only if there is a computational path $P$ of $Z$ such that $A(P) \subseteq S$.

Special classes of non-deterministic branching programs can be defined by putting restrictions
on properties of their root-leaf paths. A restriction is \emph{semantic} if it is applied to
\emph{computational} paths only and \emph{syntactic} if it is applied to all the root-leaf paths.
In order to define the restriction we use in this paper, we need an additional notation.

Let $SV$ be a permutation of variables and let $S$ be a sequence of literals of (some) 
variables occurring in $SV$. We say that $S$ is ordered \emph{according to} $SV$ if
for any two variables $X$ and $Y$ occurring in $S$, the occurrence of $X$ is ordered before 
the occurrence of $Y$ in $S$ if and only if $S$ is ordered before $Y$ in $V$.
For instance if $SV=(X_2,X_4,X_5,X_1,X_3)$ then $(\neg X_4,X_5, \neg X_3)$ is ordered
according to $SV$.

Let $P,P_1,P_2$ be paths of a directed graph $G$. Then $P=P_1+P_2$ if $P$ is obtained by 
appending $P_2$ to the end of $P_1$. For example, suppose that a path is represented by a sequence
of its vertices and let $P=(v_1,v_2,v_3,v_4,v_5)$. Then $P=(v_1,v_2,v_3)+(v_3,v_4,v_5)$
and also $P=P+(v_5)$ and also $P=(v_1)+P$. This definition is naturally extended to a a decomposition
of $P=P_1+ \dots, P_k$ into an arbitrary number of paths.

We consider a class of non-deterministic branching programs $Z$ for which there is a permutation $SV$
of its variables and a constant $c$ such that each \emph{computational} (that is, the restriction is semantic) 
path $P$ of $Z$ can be represented as $P=P_1, \dots, P_c$ so that on each $P_i$ each variable 
occurs at most once and the sequence of literals labelling the edges along $P_i$ is ordered according to $SV$.

We call this class of branching programs \emph{Nondeterministic Semantic $c$-{\sc obdd}} and abbreviate it
$c$-{\sc nsobdd}. Notice that the ordering imposed on computational paths is more restrictive than
the one imposed on read-$c$-times oblivious branching programs. Indeed, in the latter case, 
variables can occur along a path in an arbitrary (though the same for all paths) order.
In our case, however, the order of occurrences is determined by concatenation of $c$ copies of the
\emph{same} permutation of variables.

Given a {\sc cnf} $F$, its \emph{primal graph} has the set of vertices corresponding to the variables of $F$.
Two vertices are adjacent if and only if there is a clause of $F$ where the
corresponding variables both occur. 


Given a graph $G$, its \emph{tree decomposition} is a pair $(T,{\bf B})$ where $T$ 
is a tree and ${\bf B}$ is a set of bags $B(t)$ corresponding to the vertices $t$ of $T$.
Each $B(t)$ is a subset of $V(G)$ and the bags obey the rules of \emph{union} (that is, $\bigcup_{t \in V(T)} B(t)=V(G)$),
\emph{containment} (that is, for each $\{u,v\} \in E(G)$ there is $t \in V(t)$ such that $\{u,v\} \subseteq B(t)$),
and \emph{connectedness} (that is for each $u \in V(G)$, the set of all $t$ such that $u \in B(t)$ induces a subtree of $T$).
The \emph{width} of $(T,{\bf B})$ is the size of the largest bag minus one. The treewidth of $G$ is the smallest width of a tree
decomposition of $G$. 
If $T$ is a path then $(T,{\bf B})$ is a \emph{path decomposition}
The \emph{pathwidth} of a graph is the smallest width of its path decomposition. 

\begin{figure} [h]
\centering
\includegraphics[height=5cm]{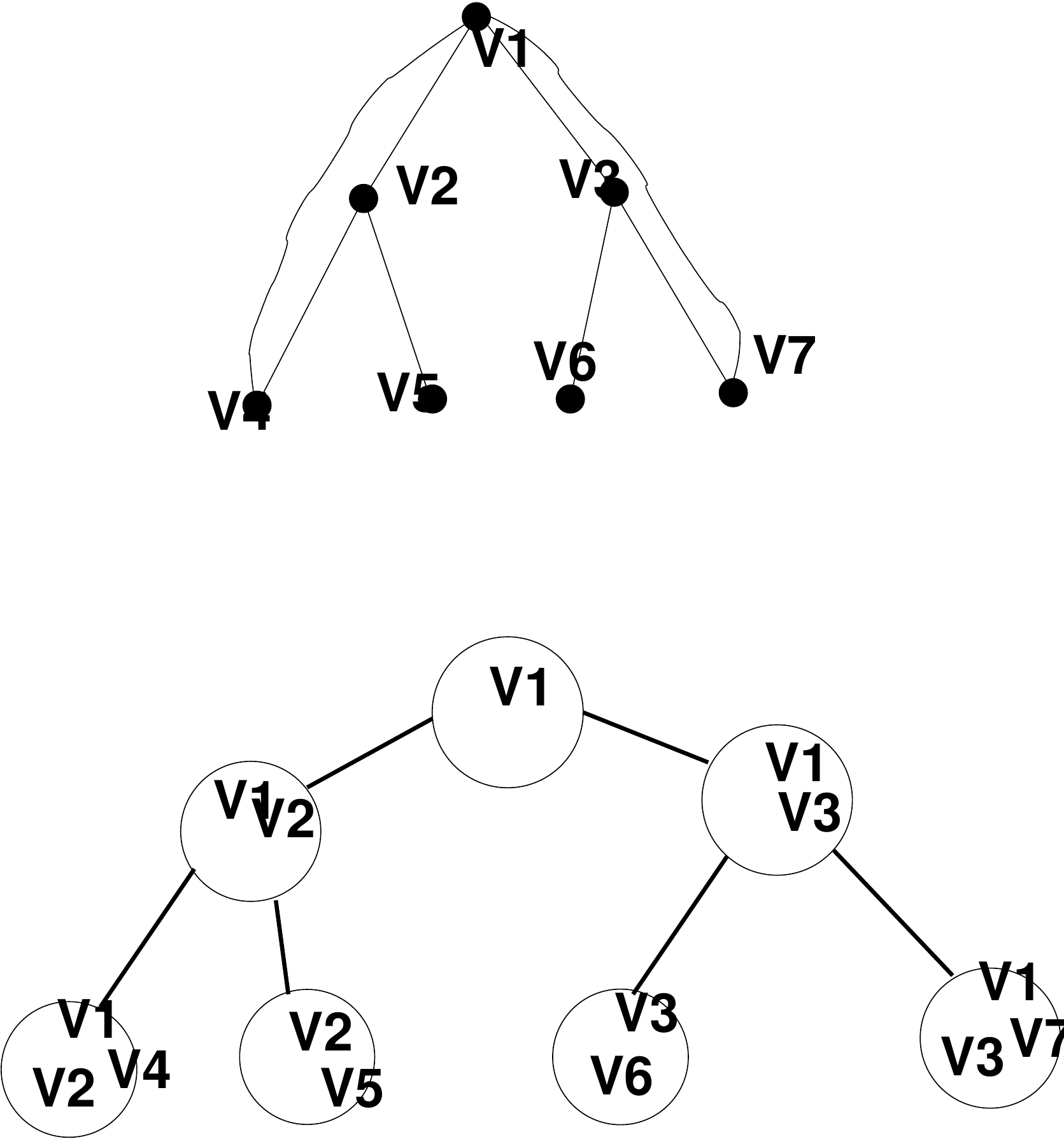}
\caption{A graph and its tree decomposition}
\label{TWDPic}
\end{figure}

Figure \ref{TWDPic} shows a graph and its tree decomposition.
The width of this tree decomposition is $2$ since the size of the largest bag
is $3$.

\section{Lower bound parameterized by treewidth} \label{lbmain}
In this section, given two integers $r$ and $k$, we define a class of {\sc cnf}s, 
roughly speaking, based on complete binary trees of height $r$ where each node is associated 
with a clique of size $k$. Then we prove that the treewidth of the primal graphs of {\sc cnf}s of this
class is linearly bounded by $k$. Further on, we state the main technical theorem (proven in the next
section) that claims that the smallest $c$-{\sc nsobdd} size for {\sc cnf}s of this class exponentially depends
on $rk$. Finally, we re-interpret this lower bound in terms of the number of variables and the treewidth to get
the lower bound announced in the Introduction.

Let $G$ be a graph. A \emph{graph based} {\sc cnf} denoted by $CNF(G)$ is defined as follows. 
The set of variables consists of variables $X_u$ for each $u \in V(G)$
and variables $X_{u,v}=X_{v,u}$ for each $\{u,v\} \in E(G)$. 
The set of clauses consists of clauses $C_{u,v}=C_{v,u}=(X_u \vee X_{u,v} \vee X_v)$
for each $\{u,v\} \in E(G)$. In other words, the variables of $CNF(G)$ correspond to the vertices
and edges of $G$. The clauses correspond to the edges of $G$. 

Denote by $T_r$ a complete binary tree of height $r$.
Let $CT_{r,k}$ be the graph obtained from $T_r$ by associating each vertex
with a clique of size $k$ and, for each edge $\{u,v\}$ of $T_r$, making all the vertices
of the cliques associated with $u$ and $v$ mutually adjacent.
Denote $CNF(CT_{r,k})$ by $F_{r,k}$. 

\begin{figure}[h]
\includegraphics[height=5cm]{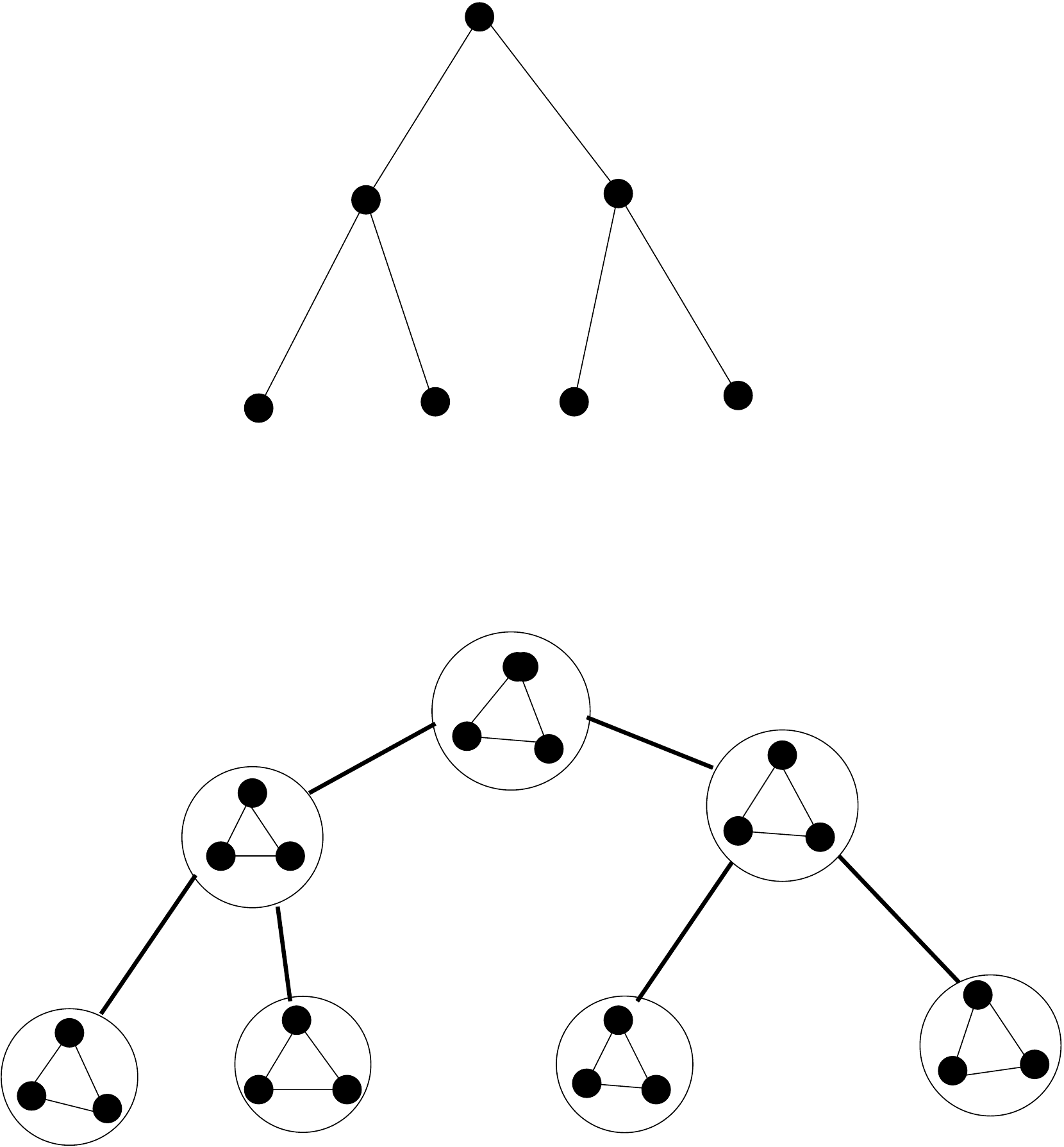}
\caption{$T_2$ and $CT_{2,3}$}
\label{TWPattern}
\end{figure}

Figure \ref{TWPattern} shows $T_2$ and $CT_{2,3}$.
To avoid shading the picture of $CT_{2,3}$ with many edges,
the cliques corresponding to the vertices of $T_2$ are marked
by circles and the bold edges between the circles mean that
that there are edges between all pairs of vertices of the corresponding 
cliques. 


\begin{lemma} \label{twf}
Let $k \geq 2$. Then
the treewidth of the primal graph of $F_{r,k}$ is at most $2k-1$. 
\end{lemma}

{\bf Proof.}
The primal graph of $F_{r,k}$ can be obtained from $CT_{r,k}$
by adding one vertex $v_e$ for each edge $e$ of $CT_{r,k}$ and
making this vertex adjacent to the ends of $e$.
Let $(T,{\bf B})$ be a tree decomposition of $CT_{r,k}$ of size
at most $2k-1$. For each vertex $v_e$, add a new vertex $x$ to $T$
adjacent to the vertex whose bag contains the ends of $e$.
Associate with $x$ a bag containing $v_e$ and the ends of $e$.
It is not hard to see that as a result we obtain a tree decomposition
of the primal graph of $F_{r,k}$. Also, as the size of each new bag
is $3$ and $k \geq 2$, the width of the tree decomposition remains at most
$2k-1$. So, it remains to show that the treewidth of $CT_{r,k}$ is at most
$2k-1$.

Consider the following tree decomposition $(T,{\bf B})$ of $CT_{r,k}$.
$T$ is just $T_r$. We look upon $T_r$ as a rooted tree, the centre of
$T_r$ being the root. The bag $B(u)$ of each node $u$ contains the clique of
$CT_{r,k}$ corresponding to $u$. In addition, if $u$ is not the root vertex
then $B(u)$ also contains the clique corresponding to the parent of $u$. 
Observe that $(T,{\bf B})$ satisfies the connectivity property.
Indeed, each vertex appears in the bag corresponding to its `own' clique and 
the cliques of its children. Clearly, the set of nodes corresponding to the bags 
induce a connected subgraph. The rest of the tree decomposition properties can
be verified straightforwardly. We conclude that $(T,{\bf B})$ is indeed a tree
decomposition of $CT_{r,k}$. As the size of each bag is at most $2k$, the width
of $(T,{\bf B})$ is at most $2k-1$.

$\blacksquare$

The following is the main technical result whose proof is given in the next section. 
\begin{theorem} \label{maintheor}
The size of {\sc obdd} computing $F_{r,k}$ is at least $2^{rk/(4c-2)}$.
\end{theorem}

The following lemma reformulates the statement of Theorem \ref{maintheor}
in terms of the number of variables of $F_{r,k}$ and $k$. 

\begin{lemma} \label{reform1}
Let $m$ be the number of variables of $F_{r,k}$.
Then the size of $c$-{\sc nsobdd} computing $F_{r,k}$ is at least
$(\frac{m}{6k^2})^{k/(4c-2)}$. 
\end{lemma}

{\bf Proof.}
Recall that $T_r$ has $2^{r+1}-1$ nodes.
For each node $a$ of $T_r$, $F_{r,k}$ has $k$ variables corresponding to the
vertices of the clique of $a$ plus ${k \choose 2}$ variables corresponding to the edges
of this clique. In addition, if $a$ is a non-root node then it is associated with $k^2$
variables connecting the clique of $a$ with the clique of its parent. Thus each node of
$T_r$ is associated with at most $k+{k \choose 2}+k^2$ variables and hence the total number of
variables $m \leq (2^{r+1}-1)*(k+{k \choose 2}+k^2) \leq 2^r*6k^2$.
Thus $2^r \geq \frac{m}{6k^2}$. 

It follows from Theorem \ref{maintheor}
that the size of a $c$-{\sc nsobdd} computing $F_{r,k}$ is at least
$2^{rk/(4c-2)}=(2^r)^{k/(4c-2)} \geq  (\frac{m}{6k^2})^{k/(4c-2)}$
as required. 
$\blacksquare$

Two lower bound parameterized by the treewdith now easily follows.

\begin{theorem}
For each $p \geq 3$ there is 
an infinite sequence of {\sc cnf}s $F_1, F_2 \dots,$ of treewidth at most $p$
of their primal graphs such that for each $F_i$ the size of $c$-{\sc obdd}
computing it is at least $(\frac{m}{3p^2})^{p/(8c-4)}$, 
where $m$ is the number of variables of $F_i$. 
In particular, for $c=1$ and every fixed $p$, we get the earlier obtained
{\sc obdd} lower bound of $\Omega(m^{p/4})$ as a special case. 
\end{theorem}

{\bf Proof.}
For an odd $p$, consider the {\sc cnf}s $F_{r,(p+1)/2}$ for all $r \geq 1$
and for an even $p$, consider the {\sc cnf}s $F_{r,p/2}$ for all $r \geq 1$.
By Lemma \ref{twf}, the treewidth of the primal graph of $F_{r,(p+1)/2}$ is at 
most $p$ and of $F_{r,p/2}$ at most $p-1$. Thus the treewidth requirement is
satisfied regarding these classes. 

Taking into account Lemma \ref{reform1} and performing simple algebraic
calculation, we observe the $c$-{\sc nsobdd} size is lower-bounded by
$(\frac{m}{3p^2})^{p/(8c-4)}$.
$\blacksquare$

Theorem \ref{maintheor} also allows us to separate between $c$-{\sc nsobdd}
and Sentential Decision Diagrams {\sc sdd}  \cite{SDD} for every fixed $c$. 

\begin{theorem}
There is an infinite family of functions that can be computed by
SDDs of size $O(n^3)$ and for which the smallest $c$-NSOBDD
are of size $n^{\Omega(\log n)}$ (for each fixed $c$) 
\end{theorem}

{\bf Proof}
Consider functions $F_{r,r}$.
Let us compute the number $n$ of variables of $F_{r,r}$. 
Following the calculation as in Lemma \ref{reform1},
we observe that
$n=(2^{r+1}-1)*(\frac{r*(r-1)}{2}+r)+(2^{r+1}-2)*r^2=
  2^{r}(3r^2+r)-\frac{5r^2+r}{2}$

Denote $3r^2+r$ by $p_1$ and $\frac{5r^2+r}{2}$ by $p_2$.
Then $r=\log \frac{n+p_2}{p_1}$. 

In particular, 
$r \geq \log n-\log p_1 \geq \log n-r$ and hence $r \geq \log n/2$ for a sufficiently 
large $r$. Substituting $\log n/2$ instead $r$ and $k$ in the lower bound provided
by Theorem \ref{maintheor} gives us lower bound $2^{\frac{\log^2 n}{16c-4}}=
n^{\frac{log}{16c-4}}$ which is $\Omega(n^{\log n}$ for every fixed $c$. 

On the other hand, for a sufficiently large $n$, $r \leq \log(n+p_2) \leq \log(2n)=\log n+1$.
By Lemma \ref{twf}, the treewidth of the primal graph of $F_{r,r}$ is at most
$2r-1$ which is at most $2\log n+1$ by the above upper bound.
Thus, according to \cite{SDD}, the size of {\sc sdd} for $T_{r,r}$ is bounded by
$O(2^{2log n}n)=O(n^3)$, as required. $\blacksquare$

\section{Lower bound parameterized by the matching width}
The central concept we use for the proof of Theorem \ref{maintheor}
is that of matching width.
A matching $M$ of a graph $G$ is a set of edges of $G$ such that no two
edges are incident to the same vertex.  
Let $SV$ be a \emph{permutation} of the set $V=V(G)$ of vertices of a graph
$G$. Let $S_1$ be a \emph{prefix} of $SV$ (i.e. all vertices of $SV \setminus S_1$ are ordered after
$S_1$). Let us call the \emph{matching width} of $S_1$,
the size of the largest matching  consisting of the edges between $S_1$ and $V \setminus S_1$
(we take the liberty to use sequences as sets, the correct use will be always clear
from the context). Further on, the matching width of $SV$ is the largest matching
width of a prefix of $SV$. Finally the \emph{matching width} of $G$, is the smallest 
matching width of a permutation of $V(G)$.

\begin{example}
Consider a path of $10$ vertices $v_1, \dots, v_{10}$ so that $v_i$ is adjacent to $v_{i+1}$ for
$1 \leq i<10$. The matching width of permutation $(v_1, \dots, v_{10})$ is $1$ since between any suffix
and prefix there is only one edge. However, the matching width of the permutation 
$(v_1,v_3,v_5,v_7,v_9,v_2,v_4,v_6,v_8,v_{10})$ is $5$ as witnessed by the partition
$\{v_1,v_3,v_5,v_7,v_9\}$ and $\{v_2,v_4,v_6,v_8,v_{10}\}$. Since the matching width of a graph is determined
by the permutation having the smallest matching width, and, since the graph has some edges, there cannot be
a permutation of matching width $0$, we conclude that the matching width of this graph is $1$.
\end{example}

The main `engine' for establishing the lower bound for Theorem \ref{maintheor} is the following theorem,
stating a lower bound on the size of a $c$-{\sc nsobdd}.

\begin{theorem} \label{cnfbound}
Let $G$ be a graph of matching width at least $t$ and let $Z$ be a $c$-{\sc nsobdd} computing $CNF(G)$. 
Then $|Z| \geq 2^{t/2c-1}$. 
\end{theorem}

Theorem \ref{cnfbound} is proved in Section \ref{engine}.
In order use Theorem \ref{cnfbound} for a proof of Theorem \ref{maintheor},
we need an additional statement providing a lower bound on 
the matching width of graphs $CT_{r,k}$
(recall $F_{r,k}=CNF(CT_{r,k})$).

\begin{theorem} \label{cltreemt} (Lemma 2 of \cite{RazgonKR14})
For any $r$, the matching width of $CT_{r,k}$ is at least $rk/2$.
\end{theorem}

Now, we are ready to prove Theorem \ref{maintheor}

{\bf Proof of Theorem \ref{maintheor}}
According to Theorem \ref{cnfbound}, the size of $Z$ implementing $F_{r,k}=CNF(CT_{r,k})$
is at least $2^{t/2c-1}$ where $t$ is the matching width of $CT_{r,k}$
Replace $t$ by the lower bound $rk/2$ on the matching width of $CT_{r,k}$
provided by Theorem \ref{cltreemt}. The required lower bound $2^{rk/(4c-2)}$
immediately follows. $\blacksquare$

\subsection{Proof of Theorem \ref{cnfbound}} \label{engine}
Let $SV$ be a permutation of the vertices of $G$ where $u$ precedes
$v$ if and only if $X_u$ precedes $X_v$ in te underlying permutation
$SV^*$ of $Z$. We refer to $SV$ as the permutation of $V(G)$ \emph{corresponding}
to $Z$. Let $SVP$ be a prefix of $SV$ such that there is a matching
$M=\{\{u_1,v_1\}, \dots, \{u_t,v_t\}\}$ of $G$ such that all of $u_1, \dots, u_t$
belong to $SVP$ and all of $v_1, \dots v_t$ do not. Such a prefix exists by definition
of matching width and our assumption that matching width of $G$ is at least $t$. 

Let ${\bf S}$ be the set of all assignments $S$ to the variables of $CNF(G)$
satisfying the following conditions.
\begin{itemize}
\item Each $\neg X_{u_i,v_i} \in S$ for $1 \leq i \leq t$.
\item For each $1 \leq i \leq t$, the occurrences of $X_{u_i}$ and
$X_{v_i}$ have distinct signs (if the former occurs positively the latter occurs
negatively and if the former occurs negatively the latter occurs positively).
\item All the variables besides $\bigcup_{i=1}^t \{X_{u_i},X_{u_i,v_i},X_{v_i}\}$ are assigned 
positively. 
\end{itemize}

Then the following statements are easy to observe.
\begin{observation} \label{sfacts}
\begin{enumerate}
\item Each $S \in {\bf S}$ is a satisfying assignment of ${\bf S}$.
\item $|{\bf S}| \geq 2^t$.
\end{enumerate}
\end{observation}

{\bf Proof.}
For the first statement, note that all the clauses $(X_u \vee X_{u,v} \vee X_v)$ besides $(X_{u_i} \vee X_{u_i,v_i} \vee X_{v_i})$ are clearly 
satisfied by $S$ because $X_{u,v}$ is assigned positively by construction. 
The clauses $(X_{u_i} \vee X_{u_i,v_i} \vee X_{v_i})$
are also satisfied by $S$ because one of $X_{u_i}$, $X_{v_i}$ is assigned positively. This proves the
first statement.

There are $2^t$ ways to assign variables $X_{u_1}, \dots, X_{u_t}$. By definition of ${\bf S}$
each such assignment can be extended to an element of ${\bf S}$ and these elements are clearly all distinct.
This proves the second statement.
$\blacksquare$

In light of the first statement of Observation \ref{sfacts}, for each $S \in {\bf S}$ we can
identify a computational path $P_S$ of $Z$ such that $A(P_S) \subseteq S$.
For each $P_S$ we are going to identify a sequence of its vertices of length at most $2c-1$
and to show that for distinct $S_1,S_2 \in {\bf S}$, the sequences associated with $P_{S_1}$ and $P_{S_2}$
are distinct. In light of the second statement of Observation \ref{sfacts}, it will follow that $Z$
contains at least $2^t$ sequences of nodes of length $2c-1$. As the number of such sequences is at most
$|Z|^{2c-1}$, it will immediately follow that $|Z| \geq 2^{t/(2c-1)}$.

In order to define a sequence of vertices associated with each $P_S$, we need some preparation.
Let $P$ be an arbitrary computational path of $Z$ and let $P_1, \dots, P_c$ be subpaths of $P$
such that $P=P_1+ \dots+ P_c$ and the following holds for each $P_i$.
\begin{itemize}
\item Each variable occurs at most once as a label of $P_i$.
\item The labels on $P_i$ are ordered according to $SV^*$.
\end{itemize}

Note that the required $P_1, \dots, P_c$ exists according to definition of $c$-{\sc nsobdd}.

Further on, let $P'_1, \dots, P'_{2c}$ be subpaths of $P$ such that
for each $1 \leq i \leq c$, the following holds.
\begin{itemize}
\item $P_i=P'_{2i-1}+P'_{2i}$.
\item For each $v \in SVP$, $X_v$ can occur only in $P'_{2i-1}$ (not in $P'_{2i}$).
\item For each $v \notin SVP$, $X_v$ can occur only in $P'_{2i}$ (not in $P'_{2i-1}$).
\end{itemize}

Note that $P'_1, \dots, P'_{2c}$ exists. Indeed, by definition of $SVP$, all the variables
$X_1=\{X_v| v \in SVP\}$ occur in $SV^*$ before all the variables $X_2=\{X_v| v \notin SVP\}$.
Therefore, if both $X_1$ and $X_2$ occur on $P_i$, we can identify the last edge $e$ of $P_i$
labelled by a variable of $X_1$ and let $P'_{2i-1}$ to be the prefix of $P_i$ ending at the head of $e$.
If only variables of $X_1$ occur on $P_i$ then let $P'_{2i-1}=P_i$ and $P_{2i}$ be the last vertex of $P_i$.
If only variables of $X_2$ occur on $P_i$ then let $P'_{2i-1}$ be the first vertex of $P_i$ and $P'_{2i}=P_i$.
Finally, if no variables occur on $P_i$, the partition can be arbitrary.

Let $x_1, \dots, x_{2c-1}$ be the respective ends of $P'_1, \dots, P'_{2c-1}$.
We call $x_1, \dots, x_{2c-1}$ \emph{the separation vector} of $P$ and $P'_1, \dots, P_{2c}$
the \emph{decomposition} of $P$ w.r.t. $x_1, \dots, x_{2c-1}$.

{\bf Remark.}
Note that there may be more than one possible $P'_1, \dots, P'_{2c}$ satisfying
the above conditions and hence $P$ can have several separation vectors.
We just pick an arbitrary one and call it \emph{the} separation vector.

The separation vectors of paths $P_S$ are these very sequences mentioned in the proof plan above.
Now we are going to prove that distinct paths $P_S$ have different separation vectors.

\begin{lemma} \label{dist1}
Let $S_1,S_2$ be two distinct elements of ${\bf S}$.
Then $P=P_{S_1}$ and $Q=P_{S_2}$ have different separation vectors.
\end{lemma}

{\bf Proof.}
Assume that $P$ and $Q$ have the same separation vector $(x_1, \dots, x_{2c-1})$.
Let $u_i$ be a variable having opposite assignments in $S_1$ and $S_2$.
(Such a variable necessarily exists because the assignments of $v_1, \dots v_t$ are
determined by assignments of $u_1, \dots, u_t$. So, if the assignments if each $u_i$ 
has the same occurrence in both $S_1$ and $S_2$, the same is true regarding each $v_i$,
and hence $S_1=S_2$, a contradiction). Assume w.l.o.g. that $u_i$ occurs negatively in 
$S_1$ and positively in $S_2$.

Let $P_1, \dots, P_{2c}$ and $Q_1, \dots, Q_{2c}$ be the respective decompositions of $P$
and $Q$ w.r.t. to $x_1, \dots, x_{2c-1}$. Let $PQ$ be the path obtained from
$P$ by replacement of each $P_j$ with even $j$ by $Q_j$. 

\begin{claim} \label{clmdiff1}
$PQ$ is a computational path.
\end{claim}

{\bf Proof.}
We need only to verify that there is no variable occurring both positively and negatively
on $PQ$. By definition of ${\bf S}$, each variable $X_{u,v}$ has the same occurrence in
both $S_1$ and $S_2$. As $A(P) \subseteq S_1$ and $A(Q) \subseteq S_2$,
$X_{u,v}$ cannot have distinct occurrences in $A(P)$ and $A(Q)$.
By definition of the decomposition w.r.t. the separation vector,
a variable $X_v$ with $v \in SVP$ cannot occur in $Q_i$ with even $i$. 
It follows that in $PQ$, $X_v$ can only occur on $P_1 \cup \dots P_{2c-1}$
which is a subgrpah of $P$. As $P$ is a computational path, it does not contain opposite literals
of $X_v$ and hence neither does $P_1 \cup \dots P_{2c-1}$. Due to the same reason, a variable $X_v$ with 
$v \notin SVP$ cannot occur on $P_i$ with even $i$. 
It follows that in $PQ$ $X_v$ can only occur on $Q_2 \cup \dots Q_{2c}$
which is a subgrpah of $Q$. As $Q$ is a computational path, it does not contain opposite literals
of $X_v$ and hence neither does $Q_2 \cup \dots Q_{2c}$.
$\square$

\begin{claim} \label{clmdiff2}
$A(PQ)$ is disjoint with $\{X_{u_i}, X_{u_i,v_i},X_{v_i}\}$.
\end{claim}

{\bf Proof.}
By definition of ${\bf S}$ $X_{u_i,v_i}$ occurs negatively in both $S_1$ and $S_2$
and hence it cannot occur positively in $A(P)$ nor in $A(Q)$ and hence, in turn
it cannot occur positively in $A(PQ)$ composed of subpaths of $P$ and $Q$. 
Since $u_i \in SVP$, a literal of $X_{u_i}$ cannot 
occur on $Q_2, \dots, Q_{2c}$ (by definition of the decomposition w.r.t. the separation vector). 
If a literal of $X_{u_i}$ occurs on $P_1, \dots, P_{2c-1}$ 
then it is an element of $S_1$ and hence negative by assumption.
Similarly, a literal of $X_{v_i}$
cannot occur on $P_1, \dots, P_{2c-1}$ (since $v_i \notin SVP$).
If a literal of $X_{v_i}$ occurs on $Q_2, \dots, Q_{2c}$ then it is an element of
$S_2$ and hence negative (by assumption, $X_{u_i} \in S_2$ and hence $\neg X_{v_i} \in S_2$
by definition of ${\bf S})$.
$\square$

By Claim \ref{clmdiff1} and definition of $Z$, an arbitrary extension of $A(PQ)$
is a satisfying assignment of $CNF(G)$. By Claim \ref{clmdiff2}, there is an extension
$S$ of $A(PQ)$ containing all of $\neg X_{u_i}, \neg X_{u_i,v_i}, \neg X_{v_i}$. 
However, $S$ falsifies clause $(X_{u_i} \vee X_{u_i,v_i} \vee X_{v_i})$
existing since $\{u_i,v_i\}$ is an edge of $G$. This contradiction shows that our
initial assumption that $P$ and $Q$ have the same separation vector is incorrect
and hence the lemma holds.
$\blacksquare$

{\bf Proof of Theorem \ref{cnfbound}}
It follows from Lemma \ref{dist1} and the second statement of Observation \ref{sfacts}
that there are at least $2^t$ distinct separation vectors of computational paths of $Z$.
Each separation vector is a sequence of nodes of $Z$. Clearly, there are at most
$Z^{2c-1}$ such sequences. That is $2^t \leq Z^{2c-1}$. Hence
$Z \geq 2^{t/(2c-1)}$, as required. $\blacksquare$

\section{Matching width vs. pathwidth}
In this section we will show that the matching width, $mw(G)$, of a graph $G$
is linearly related to its pathwidth, $pw(G)$. It particular, we will show that
$pw(G)/2 \leq mw(G) \leq pw(G)+1$.

Let us extend our notation. The maximum matching size of a graph $G$ is
denoted by $\nu(G)$. Let $SV=(v_1, \dots v_n)$ be an ordering of vertices of $G$. 
For $1 \leq i < n$, we denote $\{v_1, \dots, v_i\}$ by $V^{SV}_i$ and $V(G) \setminus V^{SV}_i$
by $\neg V^{SV}_i$. The superscript can be omitted if the ordering is clear from the context.
We denote by $G^{SV}_i$ or by $G_i$, if the ordering is clear from the context,
the graph with the set of vertices $V(G)$ and the set of edges
$\{\{u,v\}|\{u,v\} \in E(G),u \in V_i, v \in \neg V_i\}$. In other words the edges of $G_i$ are exactly those 
edges of $G$ that have
one end in $V_i$ and one end in $\neg V_i$. With this notation in mind, the matching width $mw_SV(G)$
of $SV$ can be stated as follows.

\begin{equation}
mw_{SV}(G)=max_{i=1}^n \nu(G_i)
\end{equation}

If we denote by ${\bf SV}$ the set of all permutations of vertices of $G$
then

\begin{equation}
mw(G)=min_{SV \in {\bf SV}} mw_{SV}(G)
\end{equation}

Recall that a vertex cover (VC) of graph $G$ is a set of vertices incident to all of its edges.
The smallest size of vertex cover of $G$ is denoted by $\tau(G)$. 

Observe that each $G_i$ is a bipartite graph because $V_i$ and $\neg V_i$, partitioning its set of 
vertices are indepdent sets of $G_i$. It is well known that for a bipartite graph the size of the smallest 
vertex cover equals the size of maximum matching, that is $\nu(G_i)=\tau(G_i)$. Hence $mw_{SV}(G)$
can be restated as follows

\begin{equation} \label{matvc}
mw_{SV}(G)=max_{i=1}^n \tau(G_i)
\end{equation}

Now we are bready to prove an upper bound on $mw(G)$.

\begin{theorem}
For any graph $G$, $mw(G) \leq pw(G)+1$.
\end{theorem} 

{\bf Proof.}
Let $(P,{\bf B})$ be a path decomposition of $G$ of width $pw(G)$.
Let $x_1, \dots, x_m$ be the vertices of $P$ chronologically listed as they
occur along $P$. Recall that ${\bf B}=\{B(x_1), \dots, B(x_m)\}$ are the bags
of the decomposition and the size of each bag is at most $pw(G)+1$.

Now we are going to define a permutation $SV$ of $V(G)$ for which we will
show that $mw_{SV}(G) \leq pw(G)+1$, which will imply the theorem because,
by definition $mw(G) \leq mw_{SV}(G)$. 

For $u \in V(G)$, let $f(u)$ be the smallest number $i$ such that $u \in B_{x_i}$.
Let $SV$ is an arbitrary permutation of $V(G)$ such that $u<_{SV} v$ whenever 
$f(u)<f(v)$. It is not hard to see that such an order indeed exists.
For instance, $SV$ can be created as follows. Arbitrary order the vertices of
$B(x_1)$. For each $1<i \leq n$, suppose that the vertices $B(x_1) \cup \dots \cup B(x_{i-1})$
have been already ordered and let $SV'$ be the corresponding permutation.
Then create a permuation of $B(x_1) \cup \dots \cup B(x_{i})$ by arbitrary ordering the vertices of
$B_{x_i} \setminus SV'$ and appending them to the end of $SV'$.

We are going to show that for each $1 \leq i < n$, $B(x_{f(v_i)})$ is a vertex cover of $G_i$
that is for each $\{u,v\} \in E(G_i)$ either $u \in B(x_{f(v_i)})$ 
or $u \in B(x_{f(v_i)})$.
Observe that this will imply
the desired statement that $mw_{SV}(G) \leq pw(G)+1$.
Indeed, by definition, there is $1 \leq i<n$ such that $mw_{SV}(G)=\tau(G_i)$.
Combining with the claim we are going to prove, we will have
\begin{equation}
mw_{SV}(G)=\tau(G_i) \leq B(x_{f(v_i)}) \leq pw(G)+1
\end{equation} 
the first and the second inequalities follow from the definitions of 
$\tau$ and pathwidth, respectively.

Pick $1 \leq i < n$ and let $\{u,v\} \in E(G_i)$.
Assume w.l.o.g. that $u \in V_i$ and $v \in \neg V_i$. 
Then, by definition of $SV$, $f(u) \leq f(v_i)$ and 
$f(v) \geq f(v_i)$. If the equality occurs regarding any
of them, say $f(u)=f(v_i)$ then, by definition of
function $f$, $u \in B(x_{f(u)})=B(x_{f(v_i)})$. 
Thus it remains to consider the case 
where $f(u)<f(v_i)$ and $f(v)>f(v_i)$.

By the containment property of the path decomposition,
there is $j$ such that $\{u,v\} \subseteq B(x_j)$.
By definition of $f(v)$, $f(v) \leq j$ and hence $f(v_i)<j$.
To preserve the connectedness property,
$u$ must occur in all bags $B(x_r)$ for $f(u) \leq r \leq j$.
In particular, since $f(u)<f(v_i)$ and $f(v_i)<j$,
$u \in B(x_{f(v_i)})$, as required. $\blacksquare$

Next we are going to show that $pw(G) \leq 2*mw(G)$.
For this we need the following definition.

\begin{definition} \label{defsettling}
Let $SV$ be a permutation of $V(G)$.
For each $G_i$, $1 \leq i<n$, let $VC_i$ be a smallest VC of $G_i$.
The sets $VC_1, \dots, VC_{n-1}$ are called \emph{settled} w.r.t. $SV$
if for each $1 \leq i<n-1$, $VC_i \cap \neg V_i \subseteq VC_{i+1}$
\end{definition}

The following lemma is proved in Section \ref{secsettling}.

\begin{lemma} \label{lemsettling}
For each permutation $SV$ of $V(G)$ there are $VC_1, \dots, VC_{n-1}$
that are settled w.r.t. $SV$.
\end{lemma}

Now we are ready for the theorem.
\begin{theorem}
For any graph $G$, $pw(G) \leq 2mw(G)$.
\end{theorem}

{\bf Proof.}
Let $SV=(v_1, \dots, v_n)$ be a permutation of $V(G)$ such that
$mw_{SV}(G)=mw(G)$. Let $VC_1, \dots, VC_{n-1}$ be the smallest VCs
of $G_1, \dots, G_{n-1}$, respectively, that are settled w.r.t. $SV$.

Our candidate for path decomposition of width $2mw(G)$ is a pair
$(P,{\bf B})$ where $P$ is a path $x_1, \dots, x_n$ and ${\bf B}$ is
a set of bags $B(x_1), \dots, B(x_n)$ defined as follows.
\begin{itemize}
\item $B(x_1)=VC_1 \cup \{v_1\}$.
\item For $1<i<n$, $B(x_i)=VC_{i-1} \cup VC_i \cup \{v_i\}$.
\item $B(x_n)=VC_{n-1} \cup \{v_n\}$.
\end{itemize}

In the rest of the proof we demonstrate that $(P,{\bf B})$ is indeed
a path decomposition of $G$ having width at most $2mw(G)$.
This amounts to proving the following statements. 

\begin{itemize}
\item $(P,{\bf B})$ satisfies the union property. 
      Indeed, by construction, for $1 \leq i \leq n$, $v_i \in B(x_i)$.
\item $(P,{\bf B})$ satisfies the containment property.
      Indeed, let $\{v_i,v_j\} \in E(G)$ and assume w.l.o.g. that $i<j$.
      This means that $\{v_i,v_j\}$ is an edge of each of $G_i, \dots, G_{j-1}$
      and hence each of $VC_i, \dots, VC_{j-1}$ has a non-empty intersection with
      $\{v_i,v_j\}$.

      Assume that $v_j \in VC_i$. Then, by construction, $\{v_i,v_j\} \in B(x_i)$,
      satisfying the containment property. Assume next that $v_i \in VC_{j-1}$.
      Then, by construction, $\{v_i,v_j\} \in B(x_j)$, satisfying the containment 
      property. If none of the above assumptions hold then $v_i \in VC_i$
      and $v_j \in VC_{j-1}$. It follows that there is $i \leq j'<j-1$ such that
      $v_i \in VC_{j'}$ and $v_j \in VC_{j'+1}$. Then by construction, 
      $\{v_i,v_j\} \in B(x_{j'+1})$, satisfying the containment property.
\item $(P,{\bf B})$ satisfies the connectedness property.
      Assume by contradiction that the connectedness property is violated.
      That is, there is a vertex $u$ and $i,j>i+1$ such that 
      $u \in B(x_i)$, $u \notin B(x_{i+1})$, and $u \in B(X_j)$.
      We assume that $j$ is smallest possible subject to this property, that
      is, $u \notin B(x_{j-1})$.

      Since $u \notin B(x_{i+1})$, $u \notin VC_{i}$. 
      That is $u \in B(X_i) \setminus VC_i \subseteq VC_{i-1} \cup \{v_i\}$.

      It follows that $u \in V_i$. Indeed, if $u=v_i$, this follows by definition of
      $V_i$. Otherwise, notice that since $VC_1, \dots, VC_n$ are settled,
      $VC_{i-1} \cap \neg V_i \subseteq VC_i$. As we know that $u \notin VC_i$,
      we conclude that $u \in VC_{i-1} \setminus \neg V_i=VC_{i-1} \cap V_i$

      As $u \notin VC_{j-1}$, $N_{G_{j-1}}(u) \subseteq VC_{j-1}$.
      By Definition \ref{defsettling}, $N_{G_{j-1}}(u) \cap \neg V_{j} \subseteq VC_j$.
      We claim that $N_{G_j}(u) \subseteq N_{G_{j-1}}(u) \cap \neg V_{j}$.
      This claim will imply that $N_{G_j}(u) \subseteq VC_j$ and hence $u \notin VC_j$
      by the minimality of $VC_j$  (as all the neighbours of $u$ are already there).
      This is a contradiction to our assumption, confirming correctness of the connectedness
      property. It thus remains to prove the claim.

      Let $v \in N_{G_j}(u)$. As $i<j$ and $u \in V_i$, $u \in V_j$ and hence $v \in \neg V_j$.
      Consequently, $v \in \neg V_{j-1}$. As $i<j-1$, $u \in V_{j-1}$. Thus $\{u,v\}$ is an edge
      of $G$ with one end in $V_{j-1}$, the other in $\neg V_{j-1}$. Hence $\{u,v\}$ is an
      edge of $G_{j-1}$, that is $v \in N_{G_{j-1}}(u)$ confirming the claim and the connectedness
      property as specified in the previous paragraph.

\item The width of $(P,{\bf B})$ is at most $2mw(G)$.
      That is, we have to show that for each $1 \leq i \leq n$, $|B(x_i)| \leq 2mw(G)+1$.
      By definition, $|VC_i|=\tau(G_i)$ for $1 \leq i <n$. According to (\eqref{matvc}) 
      $\tau(G_i) \leq mw_{SV}(G)$. Thus, in our case, $|VC_i|=\tau(G_i) \leq mw(G)$.
      It follows that for $1<i<n$, $|B(x_i)| \leq |VC_{i-1}|+|VC_i|+1 \leq 2mw(G)+1$. 
      Clearly, the same upper bound applies to $B(x_1)$ and $B(x_n)$.
\end{itemize}
$\blacksquare$

\subsection{Proof of Lemma \ref{lemsettling}} \label{secsettling}
Let $G=(U,V,E)$ be a bipartite graph with set of vertices $U \cup V$ and the
set of edges $E$, all having one end in $U$ the other end in $V$.
In order to prove Lemma \ref{lemsettling}, we need the following three
auxiliary statements. 

\begin{proposition} \label{auxprop1}
Let $VC'$ be a smallest VC of a $G=(U,V,E)$.
Let $X \subseteq VC'$. 
Let $VC''$ be a smallest VC of $G \setminus X$.
Then $VC'' \cup X$ is a smallest VC of $G$.
\end{proposition}

{\bf Proof.}
$VC' \setminus X$ is a VC of $G \setminus X$. Indeed, none of the edges
covered $G \setminus X$ are covered by $X$ and hence they are covered by $VC' \setminus X$.

If we assume that $VC'' \cup X$ is not a smallest VC of $G$ then $|VC'| <|VC'' \cup X|$.
That is,
$|VC' \setminus X| + |X|=|VC'|<|VC'' \cup X|=|VC''|+|X|$, from where we conclude that
$|VC' \setminus X|<|VC''|$. That is, $VC' \setminus X$ is a VC of $G \setminus X$ smaller than $VC''$ 
in contradiction to the definition of $VC''$.
$\blacksquare$

\begin{lemma} \label{auxtheor1}
Let $G=(U,V,E)$ be a bipartite graph and let $X \subseteq V$ be such that
there is $VC_1$, a smallest VC of $G$, such that $X \subseteq VC_1$.
Let $Y \subseteq V$. Then $G \setminus Y$ has a smallest VC
being a superset of $X \setminus Y$.
\end{lemma}

{\bf Proof.}
Assume that the lemma is not true. Further on, assume that
$Y$ is a largest possible subset of $V$ for which the lemma
does not hold. 

Let us represent $VC_1$ as $VC'_1 \cup VC''_1$ where $VC'_1$ consists of all
vertices of $VC_1$ incident to the edges of $G \setminus Y$ and $VC''_1$ 
consists of all vertices incident to edges of $G[U \cup Y]$. 
Denote $VC'_1 \cap VC''_1$ by $PR$. 
Observe that both $VC'_1$ and $VC''_1$ are VCs of $G \setminus Y$
and $G[U \cup Y]$, respectively. Indeed, each edge $e$ of, say $G \setminus Y$
is covered by $VC$ but it can be covered only by a vertex of $VC$ incident
to it and this vertex belongs to $VC'_1$ by definition.
Note also that $PR \subseteq U$. Indeed, an edge of $G \setminus Y$ and and edge
of $G[U \cup Y]$ cannot have a joint end that belongs to $V$.

Let $VC^*_1$ be a smallest VC of $G \setminus Y$. Observe that $PR \setminus VC^*_1 \neq \emptyset$.
Indeed, assume the opposite, that is we assume that $PR \subseteq VC^*_1$.
Note that $VC^*_1 \cup VC''_1$ is a VC of $G$ as each edge of $G$ is an edge
of either $G \setminus Y$ or of $G[U \cup Y]$. 
Note further that since $PR \subseteq VC^*_1$, $VC^*_1 \cup VC''_1=VC^*_1 \cup (VC''_1 \setminus PR)$.
Now, as $VC_1$ is a smallest VC of $G$ by definition,
$|VC'_1|+|VC''_1 \setminus PR|=|VC_1| \leq |VC^*_1 \cup (VC''_1 \setminus PR)| \leq |VC^*_1|+|VC''_1 \setminus PR|$
from where we conclude that $|VC'_1| \leq |VC^*_1|$ and hence $VC'_1$ is also a smallest VC of 
$G \setminus Y$. However, this is a contradiction because $X \setminus Y \subseteq VC'_1$.
Thus we have confirmed that $PR \setminus VC^*_1 \neq \emptyset$.

Let $Y'=N_{G \setminus Y}(PR \setminus VC^*_1)$. Note that $Y'$ is not empty as by
definition of $PR$ is element of it is incident to an edge of $G \setminus Y$.
Furthermore, as $PR \subseteq U$, $Y' \subseteq V$ and disjoint with $Y$ by definition.
That is $|Y \cup Y'|>|Y|$. By maximality of $Y$, there is a smallest VC $VC_2$
of $G \setminus (Y \cup Y')$ that includes $X \setminus (Y \cup Y')$ as a subset.
As elements of $Y'$ incident to edges of $G \setminus Y$ whose other ends are not contained
in $VC^*_1$, $Y' \subseteq VC^*_1$. Thus by Proposition \ref{auxprop1},
$VC_2 \cup Y'$ is a smallest VC of $G \setminus Y$.
However, $(X \setminus Y) =((X \setminus Y) \setminus Y') \cup ((X \setminus Y) \cap Y') \subseteq 
VC_2 \cup Y'$ . providing contradiction to our initial assumption and completing the proof.
$\blacksquare$

\begin{lemma} \label{auxtheor2}
Let $G=(U,V,E)$ be a bipartite graph.
Let $Y \subseteq V$ and let $X \subseteq U$ be such that
$X$ is a subset of a smallest VC of $G \setminus Y$.
Then $X$ is a subset of a smallest VC of $G$.
\end{lemma}

{\bf Proof.}
Let $VC_1$ be a smallest VC of $G$. If $X \subseteq VC_1$, we are
done. Otherwise, we will show that there is another smallest VC of
$G$ including $X$ as a subset. 

Given $G$,$VC_1$, and $Y$, let $VC'_1,VC''_1,PR$ be defined 
as in Lemma \ref{auxtheor1}. 

Observe that $E(G \setminus VC''_1)=E((G \setminus Y) \setminus PR)$.
Indeed, if $e \in E(G \setminus VC''_1)$ then $e \in E(G \setminus Y)$
(recall from the proof of Lemma \ref{auxtheor1} that $VC''_1$
covers all the edges of $G[U \cup Y]$. Moreover, since $PR \subseteq VC''_1$,
$e \in E((G \setminus Y) \setminus PR)$. Conversely, suppose
that $e \in E((G \setminus Y) \setminus PR)$. Then $e$ is not covered
by any vertex of $VC''_1$ (as all the vertices of $VC''_1$ covering edges
of $G \setminus Y$ belong to $PR$). Hence $e \in G \setminus VC''_1$.

Recall that $PR \subseteq U$. 
Since $G \setminus Y$ has a smallest VC including $X$, it follows from Lemma \ref{auxtheor1}
(applied to $G \setminus Y$ with playing the role of $G$ and $Y$ playing the role of $U$
 for the substitution into the statement of Lemma \ref{auxtheor1})
that $(G \setminus Y) \setminus PR$ has a smallest VC $VC_2$ including $X \setminus PR$.
Employing the previous paragraph we observe that $VC_2$ is also a smallest VC of
$G \setminus VC''_1$. By Proposition \ref{auxprop1}, $VC_2 \cup VC''_1$ is a smallest VC of $G$
including $X$ because $X=(X \setminus PR) \cup PR \subseteq VC_2 \cup VC''_1$ as required.
$\blacksquare$

{\bf Proof of Lemma \ref{lemsettling}.}
Let $VC_1$ be an arbitrary smallest VC of $G_1$. For $1 \leq i<n$, having constructed
$VC_i$, we construct $VC_{i+1}$

First we observe that $G_i \setminus v_{i+1}=G_{i+1} \setminus v_{i+1}$.
By Theorem \ref{auxtheor1}, $G_i \setminus v_{i+1}$ has a smallest VC including
$(VC_i \cap \neg V_i) \setminus \{v_{i+1}\}=VC_i \cap \neg V_{i+1}$.
Next, by Theorem \ref{auxtheor2}, $G_{i+1}$ has a smallest VC including
$VC_i \cap \neg V_{i+1}$. Set $VC_{i+1}$ to be such a smallest VC. This construction
guarantees the required property for all $i$. $\blacksquare$

\end{document}